\documentstyle[preprint,aps,prb]{revtex}
\begin{document}
\bibliographystyle{prsty}
\draft
\title{Qualitative Analysis of Causal Cosmological Models}

\author{Vicen\c c M\'endez
and Josep Triginer}
\address{Departament de F\'{\i}sica. Facultat de Ci\`encies, edifici Cc.
Universitat Aut\`onoma de Barcelona\\
E-08193 Bellaterra (Barcelona) Spain.}

\maketitle
\begin{abstract}

The Einstein's field equations of Friedmann-Robertson-Walker
universes filled with a dissipative fluid described by both the 
{\em truncated} and {\em non-truncated} causal transport equations
are analyzed using techniques from dynamical systems theory. 
The equations of state,
as well as the phase space, are different from those used in the recent 
literature. In the de Sitter expansion both the hydrodynamic
approximation and the non-thermalizing condition can be fulfilled 
simultaneously. For $\Lambda=0$ these expansions turn out to be
stable provided a certain parameter of the fluid is lower than 1/2.
The more general case $\Lambda>0$ is studied in detail as well.
\end{abstract}

\section{INTRODUCTION}
Recently, isotropic spatially homogeneous viscous 
cosmological models have been investigated using the causal (truncated and 
nontruncated)
Israel-Stewart theory of irreversible processes, to modelize the bulk
viscous transport \cite{uno}$^-$\cite{tres}. It is known that
dissipative processes may play a crucial role in the evolution of 
relativistic fluids both in cosmology and in high-energy astrophysical
phenomena. The most oftenly used theory to describe such
irreversible processes has been long since the first-order non-causal 
Eckart's
theory \cite{cuatro} which however suffers from serious pathologies and 
drawbacks, i.e., superluminal velocities and 
instabilities\cite{cinco}$^-$\cite{seis}.
In the late sixties M\"uller \cite{siete} proposed a second order theory
in which the entropy flow depended on the dissipative variables besides
the equilibrium ones. Israel and Stewart\cite{ocho}$^-$\cite{nueve}, 
and Pav\'{o}n 
{\em et al.}\cite{diez} 
developed a fully relativistic formulation on that basis, the so-called 
{\em extended} or {\em transient} thermodynamics (see ref. 11 for a 
recent
and comprehensive review of the state of the art).\par
Shortly after Israel's paper appeared, Belinskii {\em et al.}\cite{doce}
applied it to a viscous cosmological fluid using
the so-called {\em truncated} version, in which some divergence terms in the
transport equations were neglected. Most of the papers dealing with
viscous and/or heat conducting cosmological models make use of such a
truncated transport equation without stating clearly what the
implications of such a simplification may be. Recently,
some effort has been invested in analyzing to what extent the neglecting
of the divergence terms can be justified from a physical point of
view \cite{tres}$^,$\cite{trece}. As far as 
we know, Hiscock and Salmonson \cite{catorce} were the first to raise
this point in the cosmological context. These authors stressed the
key importance of the usually neglected divergence terms when
obtaining viscosity-driven inflationary solutions. However, it is now clear
that to get realistic solutions to the Einstein's field equations, 
the role played by 
the equations of state relating the different thermodynamic quantities is 
crucial. Hence the
claim in \cite{catorce} applies only to a Boltzmann gas\cite{quince}. 
In fact, the
difficulty in using the extended transport equations lies
mainly in the occurrence of some additional unknown coefficients,
whose  explicit expressions must be obtained from techniques
other than those coming from thermodynamics, either kinetic or 
fluctuation theory \cite{dseis}, more than in their intrinsic 
complexity.\par
Few exact solutions have been found to the Einstein's field equations with
a non-perfec fluid described by extended thermodynamics
\cite{dsiete}$^,$\cite{docho} (ET for short).
However, they were obtained under severe restrictions on the values for
the free parameters in the transport equations. Obviously,
any further attempt to get a deeper insight on the possible
behavior of the solutions
must rely on an approximate analysis of the equations. This
paper applies qualitative analysis
techniques to the study of causal viscous Friedmann-Robertson-Walker (FRW) 
models with and without a positive cosmological constant. It is organized as 
follows. In section II we state the
basic equations governing the models and discuss the
equations of state to be used. In section III we apply the
truncated version of ET whereas in section IV a corresponding analysis is
carried out using the {\em full} version. In both cases a null and a
positive cosmological constant are considered in turn. Section V
explores their dynamical consequences. Finally, section VI summarizes the
main conclusions of the paper.

\section{BASIC EQUATIONS}
We restrict ourselves to a FRW space-time filled with an
bulk viscous fluid and a positive cosmological constant $\Lambda$.
The stress-energy tensor is
\begin{equation}
T_{ab}=(\rho+p+\Pi)u_{a}u_{b}+(p+\Pi-\Lambda)g_{ab},
\label{eq:Tab}
\end{equation}
where $u_a$ is the four velocity, $\rho$ the energy density, $p$ the
equilibrium pressure, $\Pi$ the bulk viscous pressure. 
Einstein's field equations for the 
spatially flat case (the only one we adress in this paper) are
\begin{eqnarray}
H^2& =& \frac{\kappa}{3}\rho + \frac{\Lambda}{3},\cr
3(\dot{H} + H^2)&=& -\frac{\kappa}{2}(\rho +3 P_{eff})+\Lambda,
\label{eq:ee}
\end{eqnarray}
where $H\equiv\dot{R}/R$ is the Hubble factor, $R(t)$ the cosmic scale factor
of the Robertson-Walker metric, $P_{eff}= p + \Pi$ and $\kappa =8\pi G/c^4$. 
An upper dot denotes differentiation respect to time $t$.
We assume the fluid obeys equations of state of the form
\begin{eqnarray}
\zeta = \alpha \rho ^m , \;\; p=(\gamma -1)\rho , \;\; \tau =
\frac{\zeta}{\rho },
\label{eq:ees}
\end{eqnarray}
where  $\alpha$ is a positive constant, and $\gamma$ the adiabatic index
lying in the range $1<\gamma <2$ as the sound 
velocity $v_s/c=\gamma -1$ in the fluid must be lower
than the speed of light. $\tau(\geq 0)$ is
the relaxation time for transient bulk viscous effects, i.e. the time
the system takes in going back to equilibrium once the divergence of
the four-velocity has been switched off.
The causal evolution equation for bulk viscous pressure can be cast into
the form \cite{tres}
\begin{eqnarray}
\Pi + \tau \dot{\Pi} = -3\zeta H - \frac{b}{2}\tau \Pi \left(
3H + \frac{\dot{\tau}}{\tau} -\frac{\dot{T}}{T} - \frac{\dot{\zeta}}{\zeta}
\right),
\label{eq:mc}
\end{eqnarray}
where $b =0$ for the truncated theory and $b =1$ for the full 
one. Since a dissipative
expansion is non-thermalizing, the relaxation time must exceed the 
expansion rate $H^{-1}$. This leads to
\begin{eqnarray}
\tau ^{-1} <H
\label{eq:tau}
\end{eqnarray}
which is a condition that reduces the interval of
values of $\gamma$ for which the model holds. As we shall see this restriction
may be violated in the truncated theory as well 
as in the full theory when an ideal gas equation of state is assumed. This
conflict can be circumvented by resorting to the expression for the
speed of the viscous signals $a\equiv v^2/c^2\sim\zeta/\tau\rho$, which roughly
implies \cite{tres}$^,$\cite{trece}
\begin{eqnarray}
\tau = \frac{\zeta}{a\rho},\ \ \ 0<a<1,
\label{eq:tau1}
\end{eqnarray}
\noindent and using (\ref{eq:tau1}) instead of (\ref{eq:ees}c).\par
Most of the stability analysis below
will be carried out for the de Sitter solutions, $H =$ const. As the
universe undergoes a de Sitter expansion it could be argued that the
hydrodynamic description (absolutely needed in our study if the
results are to hold) might break down. In order the fluid approach
remains valid the mean
collision time $t_{col}$ must be less (in fact much less) than the expansion 
rate, i.e., $t_{col}<H^{-1}$. From kinetic
theory one has $t_{col}=1/n\sigma$ where $n$ is the particle number density and
$\sigma$ the cross section for collisions. In general $\sigma(T)$ is an 
increasing function of the temperature whereas, for a FRW universe, $n\propto
1/R^3(=e^{-3H_{0}t}$ for a de Sitter universe). From equations
(\ref{eq:ee}a), (\ref{eq:ees}a) and (\ref{eq:tau1}) the non-thermalizing 
condition (\ref{eq:tau}) and the condition for the hydrodynamic approximation 
imply
\begin{equation}
\frac{e^{3H_{0}t}}{n_{0}\sigma}<H^{-1}_{0}<\frac{\alpha}{a}\left(\frac{3}
{\kappa}\right)^{m-1}H_{0}^{2m-2},
\label{eq:hydr}
\end{equation}
\noindent where $n_{0}$ is a positive but otherwise arbitrary 
integration constant. 
Later it will be shown that the second inequality in (\ref{eq:hydr})
can be fulfilled when suitable values for the arbitrary parameters are chosen.
Moreover, the first inequality may hold for sufficiently early times (when
the inflation era supposedly took place). As the temperature remains constant
during this period the cross section $\sigma$ can be taken approximately 
constant\cite{dnueve} .

Recently in performing the qualitative analysis
of imperfect fluid cosmological models (see for instance ref. 1, 2, 20,
21) 
dimensionless equations of state were used in terms of the dimensionless 
variables $x$ and $y$ defined as 
\begin{equation}
x\equiv 3\rho/\Theta^2,\ \ \ \ y\equiv 9\Pi/\Theta^2.
\label{eq:variables}
\end{equation}
The equations of state (eq.(3.4a,b) in ref. 20)
\begin{equation}
p/\Theta^2=p_{0}x^l,\ \ \zeta/\Theta=\zeta_{0}x^m, 
\label{eq:coleyeq}
\end{equation}

\noindent with $\Theta(\equiv 3H)$ the expansion factor,
coincide with (\ref{eq:ees}a,b) only
for $l=m=1/2$. Furthermore, for the spatially flat
FRW metric with $\Lambda =0$, the case we are interested in, we have $x=1$.
Then the bulk viscous coefficient $\zeta$ varies as $\Theta$ 
irrespective of $m$, which restricts (\ref{eq:coleyeq}) to just one case: 
$m=1/2$ in (\ref{eq:ees}a). In this paper we shall 
consider only the {\em spatially flat} case ($k=0$) which allows us
to take $(\dot{H}, H)$ as suitable dynamical variables in the phase space.
In this case it appears to be more natural, especially
when the above comments are taking into account, to adopt the
oftenly used equations of state (\ref{eq:ees}) rather than (\ref{eq:coleyeq})
in order to be able to compare our results with those in the literature for
$m\not = 1/2$. Consequently, all the fixed points to be analyzed
will correspond to either de Sitter or static spacetimes
$(X\equiv \dot{H}=0)$ being the former physically relevant in
inflationary models. If one is interested in studying non-flat FRW models the 
variables $(\dot{H}, H)$ become no longer suitable. In such a case 
those proposed by Coley \cite{veinte} are more convenient.

\section{QUALITATIVE ANALYSIS USING THE TRUNCATED THEORY} 
From equations (\ref{eq:ee}), (\ref{eq:ees}) and the expression 
(\ref{eq:tau1}) for $\tau$ we find for the Hubble factor the equation
\begin{eqnarray}
&&\ddot{H} + 3\gamma H\dot{H} +\frac{a}{\delta}\left(
H^2-\frac{\Lambda}{3}\right)^{1-m} \dot{H}\cr
&&+\frac{(3H^2-\Lambda)a}{2}\left[
\frac{\gamma}{\delta} \left( H^2-\frac{\Lambda}{3}\right)^{1-m}-3H\right ] =0.
\label{eq:hfcc1}
\end{eqnarray} 

where $\delta = \alpha (3/ \kappa)^{m-1}$. Equation (\ref{eq:hfcc1})
can be recast into the form 
\begin{eqnarray}
\dot{H}& =& P(H,X) \cr
\dot{X}&=& Q(H,X)
\label{eq:sis}
\end{eqnarray}
where
\begin{eqnarray}
P(H,X)&=&X,\label{eq:PQ1}\\
Q(H,X)&=&-3\gamma HX-\frac{a}{\delta}\left(H^2-\frac{\Lambda}{3}\right)^{1-m}
X\nonumber\\&&-\frac{(3H^2-\Lambda)a}{2}\left[\frac{\gamma}{\delta}\left(H^2-
\frac{\Lambda}{3}\right)^{1-m}-3H\right].
\label{eq:PQ2}
\end{eqnarray}
The qualitative analysis begins by linearizing the system
(\ref{eq:sis}) for small perturbations - where the linear theory holds. 
Then the Jacobian matrix
\begin{eqnarray}
\mbox{\boldmath $L$} = \left( \begin{array}{cc} P_H & P_X \\ Q_H & Q_X
\end{array} \right),\ \ \ \ \mbox{with}\ P_H\equiv \partial P/\partial 
H,\ \mbox{etc.}, 
\label{eq:L}
\end{eqnarray}
can be constructed. The 
elements of this matrix
must be evaluated at the equilibrium points $(h_i,X_i)$ (de
Sitter and static solutions) which
are found by solving the system $P(h_i,X_i) = Q(h_i,X_i)=0$.
After diagonalizing $\mbox{\boldmath $L$}$
and obtaining its eigenvalues we can decide about the type of fixed points and
their stability.\par
The analysis of the system (\ref{eq:PQ1}), (\ref{eq:PQ2}) for the two
cases with $\Lambda =0$ and $\Lambda>0$ will be carried out in turn.\\

\noindent\underline{i. $\Lambda =0$}\par

$\bullet$ $m=1/2$\\
We have the trivial fixed point
\begin{eqnarray*}
(0,0),
\end{eqnarray*}
which corresponds to an unstable static model. However, this case does not
make sense as, by Einstein's equation (\ref{eq:ee}a), $\rho=0$. If $\gamma$ 
and $\delta$ fulfill the restriction
\begin{eqnarray}
\frac{\gamma}{\delta}=3,
\label{eq:pep1}
\end{eqnarray}
there exists an infinity of fixed points $(h_{0},0)$, where $h_{0}$
denotes an arbitrary positive real constant. In this case the fixed points are 
parallel stable straight lines.\par

$\bullet$ $m\neq 1/2$\\ 
In the intervals $0\leq\ m < 1/2$, $1/2 < m <2$ there are two fixed points
\begin{eqnarray}
(0,0), \;\;\; (h_0 , 0),
\label{eq:pf1}
\end{eqnarray}
with
\begin{eqnarray}
h_0 = \left( \frac{3\delta}{\gamma}\right)^{\frac{1}{1-2m}},
\label{eq:h0}
\end{eqnarray}
whereas for $m\geq 2$ there is only one fixed point $(h_{0},0)$. The discussion
for the point $(0,0)$ mimics that for $m=1/2$.\par
Let us define the auxiliar parameter
\[
\Sigma _1 = \frac{1}{4a}\left(\gamma+\frac{a}{\gamma} \right)^2.
\] 
For $m<\frac{1}{2}-\Sigma _1$ the equilibrium point
$(h_0,0)$ is an asymptotically stable 
focus, for $m=\frac{1}{2}-\Sigma _1$ it is an asymptotically
stable degenerate node, whereas for $ m>\frac{1}{2}-\Sigma _1$ two 
cases arise. If 
$\frac{1}{2}- \Sigma _1<m<\frac{1}{2}$, then
the equilibrium point is an asymptotically stable node, whereas if 
$m>1/2$ it is a unstable saddle point.\par
In the paper by Pav\'on {\em et al}\cite{vdos} sligthly different
techniques were used to analyze the case $m\not = 1/2$ and $a=1$. Their
relevant parameter was our $\Sigma_1$ with $a=1$. Our results agree
with those of the mentioned reference (see \S 3.1 of ref. 22) providing a 
more accurate classification of the stability points.\\

\noindent\underline{ii. $\Lambda>0$}\\

As we shall see, there are two fixed points: $(h^{\Lambda}_{0},0)$ and 
$(h^{\Lambda}_{1},0)$.
From (\ref{eq:hfcc1}) it follows the equation for the fixed points 
\begin{eqnarray}
(3h_i^2-\Lambda)\left[\frac{\gamma}{\delta}\left(h_i ^2 - \frac{\Lambda}{3}
\right)^{1-m}-3h_i\right ] =0,
\label{eq:pf}
\end{eqnarray} 
which must be solved for different values of $m$. However, (\ref{eq:pf}) has 
an obvious 
solution independent of $m$, $h^{\Lambda}_{0} = \sqrt{\Lambda /3}$, which can 
be shown 
to correspond to a saddle point. This solution will be ruled
out however since it would imply that the energy density vanishes identically.
The other solution $h^{\Lambda}_1$ will be analyzed for 
$m=0, \frac{1}{2}, 1$, in turn.\par

$\bullet$ $m=0$\\
Setting to zero the big square parenthesis in (\ref{eq:pf}) and
solving the resulting equation one obtains
\[
h^{\Lambda}_1 = \frac{3\delta}{2\gamma}+\frac{1}{2}\sqrt{\frac{9\delta
^2}{\gamma ^2}+\frac{4\Lambda}{3}}.
\]
We define 
\[
 \Sigma^{\Lambda}_1 = \frac{2-2\Sigma _1}{2\Sigma _1 -1} ,\  
\Lambda _0 = \frac{27\delta ^2}{\gamma ^2} \frac{1+\Sigma^{\Lambda}_1}
{(\Sigma^{\Lambda}_1)^2}.
\] 
For $\Lambda <\Lambda _0$ the fixed point $(h^{\Lambda}_1,0)$ is an 
asymptotically stable node. If $\Lambda = \Lambda _0$ the
fixed point is an asymptotically stable degenerate node whereas for $\Lambda >
\Lambda _0$ it is an asymptotically stable focus.\par

$\bullet$ $m=1/2$\\
Now $h^{\Lambda}_1$ is given by
\[
h^{\Lambda}_1 = \sqrt{\frac{\Lambda /3}{1-\frac{9\delta ^2}{\gamma ^2}}}.
\]
For $0<9\delta ^2/\gamma ^2<1$, $(h^{\Lambda}_1, 0)$ is an
asymptotically stable node  whereas for $9\delta ^2/\gamma ^2 >1$ the
fixed point is an asymptotically stable focus for any $\Lambda>0$.\par

$\bullet$ $m=1$\\
Now $h^{\Lambda}_1 = \gamma/3\delta$. 
For $\Lambda < \gamma ^2/3\delta ^2$ we have a saddle fixed
point, whereas for $\gamma^2/3\delta^2<\Lambda<\Lambda_{1}$, where
\[
\Lambda _1 = \frac{\gamma ^2}{3\delta ^2}\Sigma^{\Lambda}_2 >0 \ \
\mbox{with}\ \ \Sigma ^{\Lambda}_2 = \frac{1}{2a}\left(\gamma +\frac{a}
{\gamma} 
\right)^2+1,
\]
the fixed point is an asymptotically stable node. For 
$\Lambda = \Lambda _1$, it is an
asymptotically stable degenerate node whereas for $\Lambda>\Lambda _1$,
it is an asymptotically stable focus.\par

\section{QUALITATIVE ANALYSIS USING THE FULL THEORY}
Actually, a proper study of viscous phenomena in the frame of ET requires 
the use of the full equation (\ref{eq:mc}) (i.e. $b=1$). The physical
implications of neglecting the second term of (\ref{eq:mc}) have been
analyzed in detail in ref. 3, 13. The use of 
(\ref{eq:mc})
requires an explicit expression for the temperature $T$ in terms of
other variables such as $\rho$ and/or $n$. So far
(with the exception of\cite{catorce}) the expression adopted for $T$ 
has been a power-law
\begin{equation}
T=\beta\rho^r,
\label{eq:tmar}
\end{equation}
\noindent where $r\geq 0$ and $\beta > 0$ are constants, which is the simplest
way to guarantee a positive heat capacity\cite{vtres}. However,
we shall see that standard thermodynamic relations restrict the
range of $r$. C\~{a}lvao {\em et al.}\cite{vcuatro} found a general equation 
for the evolution of temperature when two equations of state 
\begin{eqnarray}
\rho&=&\rho(T,n),\cr
p&=&p(T,n),
\label{eq:rhop}
\end{eqnarray}
\noindent are given. However, their equation was obtained in the
context of matter creation where $\Pi$
is reinterpreted as a non-equilibrium pressure associated to particle
production. The same equation has been carefully analyzed in \cite{trece},
it reads
\begin{equation}
\frac{\dot T}{T}=-\Theta\left[\frac{(\partial p/\partial
T)_{n}}{(\partial\rho/\partial T)_{n}}+\frac{\Pi}
{T(\partial\rho/\partial T)_{n}}\right].
\label{eq:zimeq}
\end{equation}
Obviously, when the equations of state (\ref{eq:rhop}) are known the
evolution of $T$ is no longer free but fixed by (\ref{eq:zimeq}).
However, only in very few cases these equations are explicitly known,
as for instance in the case of a radiation gas or an ideal
gas\cite{vcinco} .
Equation (\ref{eq:tmar})
generalizes in a simple way the Stefan-Boltzmann ($r=1/4$)
equation which holds for a radiation-dominated fluid in equilibrium.
Thus, in that case we get that both equations of state $\rho$ and $p$ 
(when a 
$\gamma$-law is used) have $T$ as the only independent variable, i.e.,
$\partial\rho(p)/\partial T=d\rho(p)/dT$. When the integrability
conditions for the Gibbs equation are considered an useful and interesting
relation follows \cite{vseis}

\begin{equation}
\left(\frac{\partial\rho}{\partial n}\right)_{T}=\frac{\rho+p}{n}-
\frac{T}{n}\left(\frac{\partial p}{\partial T}\right)_{n},
\label{eq:weinberg}
\end{equation}
\noindent which, by virtue of (\ref{eq:ees}b) and (\ref{eq:tmar}), yields
\begin{equation}
r=\frac{\gamma-1}{\gamma}\ (\Rightarrow 0<r<1/2),
\label{eq:rg}
\end{equation}
\noindent i.e., $r$ is no longer an independent parameter (we are indebted to 
Roy Maartens for pointing us out this restriction). It has been
argued\cite{dos} that the inequality $r<1$ is reasonable from a physical 
point of view, since ultrarelativistic and cold non-relativistic
matter have $r=1/4$ and $r\sim 2/3$, respectively.\par
However, an alternative equation can be used for $T$ instead of 
(\ref{eq:tmar}). It is well-known that a relativistic ideal monoatomic gas is 
described by the
two equations of state $p=nT$ and $\rho=3nT+m^2M$, where
$m$ is the mass of the particles and $M$ the zero-order moment of the 
Maxwell-Boltzmann distribution function (we use units $k_B=1$, $k_B$ being
the Boltzmann constant). We see that the $\gamma$-law ($\gamma$ constant) is 
{\em not} compatible with the equations of state of a monoatomic gas in
{\em equilibrium} except for radiation ($m=0$). In that case we have $n\propto 
T^{3}$ and the two equations of 
state for $p$ and $\rho$ reduce to the Stefan-Boltzmann equation and the
$\gamma$-law with $\gamma=4/3$.\par
In the remainder of this section the full viscous transport equation
will be analyzed resorting to the two expressions for the temperature 
mentioned above: a power-law given by (\ref{eq:tmar}) and an ideal gas 
equation for $p$ together with the $\gamma$-law defining $\rho$, i.e.
\begin{equation}
p=nT,\ \ \ \rho=\frac{nT}{\gamma-1}.
\label{eq:idealeq}
\end{equation}
Note that now both $T$ and $n$ are independent variables and only in
the equilibrium limit the particle number density depends exclusively on the
temperature, $n=n(T)$ (see comments above). It must be stressed that
the Stefan-Boltzmann equation together with an ideal gas equation of
state (with $n\propto T^{3}$) implies $\Pi=0$. So we conclude that out of 
equilibrium we are forced to adopt one of the two possibilities: (i) a 
power-law for $T$ with no dependence on $n$ at all; (ii) an ideal gas
equation of state for the pressure together a $\gamma$-law, with $n$
an independent variable on the same footing as $T$. Both
approaches will be considered in turn.

\subsection*{A. Potential law for the temperature}  

Using equations (\ref{eq:ee}), 
(\ref{eq:ees}), (\ref{eq:tau1}), (\ref{eq:tmar}) and (\ref{eq:rg}) the 
equation governing the evolution of the Hubble factor reduces to
\begin{eqnarray}
&&\ddot{H} + \frac{3}{2}[1+\gamma(1-r)]H\dot{H} +\frac{a}{\delta}\left(
H^2-\frac{\Lambda}{3}\right)^{1-m} \dot{H}\cr
&&+\frac{(3H^2-\Lambda)a}{2}\left[
\frac{\gamma}{\delta} \left(
H^2-\frac{\Lambda}{3}\right)^{1-m}-3(1-\frac{\gamma}{2}) H\right ]\cr
&&-3(r+1)\frac{H\dot{H} ^2}{3H^2-\Lambda}=0.
\label{eq:H2}
\end{eqnarray}

\noindent\underline{i. $\Lambda=0$}\par

$\bullet$ $m=1/2$\\
As in Section III only the case with $\gamma$ and $\delta$ fulfilling the 
restriction
\begin{eqnarray}
\frac{\gamma}{\delta}= \frac{3}{2}(2-\gamma),
\label{eq:fulfil}
\end{eqnarray}
is physically meaningful. In such instance there exists an infinity of stable 
fixed points $(h_1,0)$, with
$h_{1}$ an arbitrary positive real number. The phase portrait are
parallel stable straight lines.\par

$\bullet$ $m\neq 1/2$\\
There are two fixed points, $(0,0)$ and $(h_1,0)$, in the
interval $m \in [0,\frac{1}{2}) \bigcup (\frac{1}{2},2)$ where 
\begin{eqnarray}
h_1 =\left( \frac{3\delta(2-\gamma)}{2\gamma}\right)^{
\frac{1}{1-2m}}.
\label{eq:h1m12}
\end{eqnarray}
For $m\geq 2$ only the $(h_1,0)$ fixed point occurs, which we
analyze next as nothing new arises about the point $(0,0)$.\par
Let us define the parameter
\[
\Sigma _2 =\frac{[\gamma (2-a)+2a]^2}
{8a\gamma ^2 (2-\gamma)}>0.
\]

For $m<\frac{1}{2}-\Sigma _2 $ the
equilibrium point is an attractor in the phase space (asymptotically 
stable focus). For $\frac{1}{2}-\Sigma_2\leq m<\frac{1}{2}$ 
we have asymptotically stable nodes instead. Finally if
$m>1/2$, the fixed point is a saddle.\\

\noindent\underline{ii. $\Lambda>0$}\\

From eq.(\ref{eq:H2}) it follows the equation for the fixed points
\begin{eqnarray}
(3h_i^2-\Lambda)\left[\frac{\gamma}{\delta}\left(h_i ^2 - \frac{\Lambda}{3}
\right)^{1-m}-3(1-\frac{\gamma}{2})h_i\right ] =0,
\label{eq:pfpl23}
\end{eqnarray}
\noindent where, as in the truncated case, only the solutions vanishing the 
big square parenthesis make sense from a physical point of view. 
Eq.(\ref{eq:pfpl23})
will be solved only for three different values of $m$. In this case we must 
take into account the constraint (\ref{eq:rg}).\par

$\bullet$ $m=0$\\
The fixed point is $(h^{\Lambda}_2,0)$ with
\[
h^{\Lambda}_2 =  \frac{3\delta}{2\gamma}(1-\frac{\gamma}{2}) +\frac{1}{2}
\sqrt{\frac{9\delta ^2}{\gamma
^2}(1-\frac{\gamma}{2})^2+\frac{4\Lambda}{3}}.
\] 
Defining the two new parameters
\[
\Sigma^{\Lambda}_3 =
\frac{2(1-\frac{\gamma}{2})}{\frac{1}{a}[1-\frac{a}{2}
+\frac{a}{\gamma}]^2-2(1-\frac{\gamma}{2})}-1, 
\]
and
\[
\Lambda _2 = \frac{27\delta ^2}{\gamma ^2}(1-\frac{\gamma}{2})^2
\frac{1+\Sigma^{\Lambda}_3}{(\Sigma^{\Lambda}_3)^2},
\]  
we see that for $\Lambda\leq\Lambda _2 $ the fixed point is an
asymptotically stable node, whereas for $\Lambda >\Lambda _2 $ it is 
an asymptotically stable focus.\par

$\bullet$ $m=1/2$\\
The fixed point is
\[
h^{\Lambda}_2= \sqrt{\frac{\Lambda /3}{1-\frac{9\delta ^2}{\gamma
^2}(1-\frac{\gamma}{2})^2}},
\]
so 
\[
\delta<\frac{2\gamma}{3(2-\gamma)}.
\] 
Let us introduce 
\[
\delta _0 = \frac{2\gamma}{3(2-\gamma)}(1-\Sigma _2)^{1/2}.
\]
For $\delta >\delta _0$, $(h^{\Lambda}_2,0)$ is found to be an asymptotically 
stable node,
however if $\delta = \delta _0$ it is an asymptotically stable
degenerate node,
whereas for $\delta <\delta _0$ the fixed point is an asymptotically
stable focus. In the radiation case - i.e. $\gamma=4/3$- $\delta _0 <0$ and 
the fixed point is a stable node.\par

$\bullet$ $m=1$\\
Here 
\[
h^{\Lambda}_2 = \frac{\gamma}{3\delta (1-\frac{\gamma}{2})}.
\]
Let us define the parameter
\[
\Lambda _3 =\frac{\gamma ^2(1+2\Sigma _2)}
{3\delta ^2(1-\frac{\gamma}{2})^2}>0.
\]
\par
For $\Lambda >\Lambda _3$ the fixed point is an asymptotically stable focus, 
and for 
$\Lambda =\Lambda _3$ an asymptotically stable degenerate node.\par
Finally when $\Lambda <\Lambda _3$ we can distinguish two subcases.
Defining 
\[
\Lambda _3 ^*= \frac{\gamma ^2}{3\delta^2(1-\frac{\gamma}{2})^2},
\]
we have a saddle point for 
\[\Lambda <\Lambda _3 ^*,
\]
and an asymptotically stable node for
\[
\Lambda_3 ^*<\Lambda<\Lambda _3.
\]

\subsection*{B. Ideal gas equation for the temperature}
In this subsection we shall study the specific behavior of the equilibrium 
points making use of the state equations (\ref{eq:idealeq}). As  
the number of particles is conserved $n$ obeys the 
conservation equation
\begin{eqnarray}
\dot{n} + 3Hn =0,
\label{eq:n}
\end{eqnarray}
which leads to $n \propto R^{-3}$. The expression for the temperature 
\begin{eqnarray}
T = \frac{3}{\kappa}\frac{\gamma -1}{n_0}R^3\left(H^2-\frac{\Lambda}{3}\right),
\label{eq:temp}
\end{eqnarray}
where $n_0>0$ is a constant, follows easily. Using
(\ref{eq:ee}), (\ref{eq:ees}a,b), (\ref{eq:mc}), (\ref{eq:tau1}) and 
(\ref{eq:temp}) we get, for the evolution of the Hubble factor, the equation
\begin{eqnarray}
&&\ddot{H} + \frac{a}{\delta}\left(
H^2-\frac{\Lambda}{3}\right)^{1-m} \dot{H}+\frac{(3H^2-\Lambda)a}{2}\left[
\frac{\gamma}{\delta} \left(
H^2-\frac{\Lambda}{3}\right)^{1-m}-3(1-\frac{\gamma}{2}) H\right]\cr 
&&-6\frac{H\dot{H} ^2}{3H^2-\Lambda}=0.
\label{eq:hfccig1}
\end{eqnarray}

\noindent\underline{i. $\Lambda=0$}\\

We have the same fixed points 
as in the truncated theory (see equation (\ref{eq:h0})).\par
For $m=1/2$ the discussion parallels that of the 
truncated theory.
After linearizing the system and introducing the parameter
\[
\Sigma _3 =\frac{a}{4\gamma ^2},
\]
the following will be discussed. For $m<\frac{1}{2}-\Sigma _3$ the eigenvalues 
are complex and the equilibrium point is an attractor (asymptotically stable 
focus). For $m=\frac{1}{2}-\Sigma _3$ there is a bifurcation 
point which is an asymptotically stable degenerate node. For $m>1/2$
one has a saddle point. Finally, if 
$\frac{1}{2}-\Sigma _3<m<1/2$ the fixed point results an asymptotically 
stable node.\\

\noindent\underline{ii. $\Lambda>0$}\\

Now the fixed points $(h^{\Lambda}_2,0)$ are again the same as in the full 
theory using a power law for the temperature.\par

$\bullet$ $m=0$\\
For any $\Lambda>0$ the fixed point is an asymptotically stable focus.\par

$\bullet$ $m=1/2$\\
Defining 
\[
\delta _1 = \frac{\gamma}{3}\sqrt{1-\frac{1}{2\gamma ^2}}
\]
we note that if $\delta$ lies in the interval $0<\delta<\delta _1$
the fixed point is an asymptotically stable focus. If $\delta = \delta _1$ 
it is an asymptotically stable degenerate node, and if
$\delta_1<\delta<\gamma/3$ an asymptotically stable node for any 
$\Lambda>0$.\par

$\bullet$ $m=1$\\
Let us define
\[
\Lambda_4 = \frac{\gamma ^2}{3\delta ^2}(1+2\Sigma _3).
\] 
If $\Lambda<\gamma / 3\delta ^2 $
then the fixed point is a saddle point, but if $\gamma / 3\delta ^2<\Lambda
<\Lambda_4$ it is an asymptotically stable node. For
$\Lambda = \Lambda_4$ it is an asymptotically stable degenerate node,
and for $\Lambda >\Lambda_4$ an asymptotically stable focus.\\

{\bf Non-thermalizing condition for dissipative de Sitter expansion}\\

\noindent\underline{i. $\Lambda=0$}\\
From (\ref{eq:tau}) and (\ref{eq:ee}) one finds
\begin{eqnarray}
H^{1-2m}<\frac{\delta}{a}.
\label{eq:terc}
\end{eqnarray}\par
For the truncated and full theory using an ideal gas equation for $T$
this condition reduces to $\gamma >3a$ by virtue of (\ref{eq:h0}). 
On the
other hand, as the velocity of the viscous pulses, as well as the speed of 
sound, cannot exceeds the speed of light ($1<\gamma <2$) we obtain the 
restrictions on $\gamma$ and $a$. If $a$ lies in the range $0<a<\frac{1}{3}$, 
the two mentioned conditions amount to $1<\gamma<2$; whereas if 
$\frac{1}{3}<a<\frac{2}{3}$ 
these restrictions imply $3a<\gamma<2$. Finally, if 
$\frac{2}{3}<a<1$, no $\gamma$ can fulfill both conditions.\par
For the full theory with a power law for temperature one obtains the 
restriction
$\gamma >\gamma _c$ where 
\[
\gamma _c = \frac{6a}{3a+2}.
\]
$\gamma$ must meet simultaneously two conditions: $1<\gamma<2$ and 
$\gamma >\gamma_c$. For $0<a<\frac{2}{3}$, these restrictions imply
$1<\gamma<2$ (since $\gamma_{c}<1$); whereas for $\frac{2}{3}<a<1$
one has $1<\gamma _c<\gamma<2$. 
So for $a<1$ the full theory with a power law for
temperature always holds.\\

\noindent\underline{ii. $\Lambda>0$}\\

Instead of (\ref{eq:terc}) we now have
\begin{eqnarray}
\frac{a}{\delta}\left(H^2-\frac{\Lambda}{3} \right)^{1-m} <H
\label{eq:terc1}
\end{eqnarray}
when a positive cosmological constant is present.\par
For the fixed point 
$(h^{\Lambda}_1,0)$ (that of the truncated theory) the
restrictions for $\gamma$ are the same that in the truncated case
with $\Lambda=0$, whereas for the full theory the restriction 
(\ref{eq:terc1}), when applied to the point $(h^{\Lambda}_2,0)$, coincides
with that of the full theory using a power law for the temperature
with vanishing $\Lambda$.\par

\section{DYNAMICAL CONSEQUENCES}
In this section we study the dynamical implications 
of linearizing the equation for $H$. This
linearitzation allows one to obtain an analytical solution for $R(t)$ near the 
equilibrium points. The matrix 
\mbox{\boldmath $L$} is given by (\ref{eq:L}) and the system of
differential equations to solve is
\begin{eqnarray}
\left( \begin{array}{c} \dot{h} \\ \dot{X} \end{array} \right) =
 \left( \begin{array}{cc} P_H & P_X \\ Q_H & Q_X
\end{array} \right) \left( \begin{array}{c} h \\ X \end{array} \right)
\label{eq:sts}
\end{eqnarray}
where $h\equiv H-h_i$ being $(h_i,X_i)$ the fixed points.
Equation (\ref{eq:sts}) can be written as
\begin{eqnarray}
\ddot{h}-Q_X \dot{h} -Q_H h =0,
\label{eq:equa}
\end{eqnarray}
where in our model $P_H =0$ and $P_X =1$.
The corresponding characteristic equation reads
\[
\lambda{\pm} = \frac{Q_X \pm \sqrt{Q_X ^2 +4Q_H}}{2} 
\]
which coincides
with the equation for the eigenvalues of $\mbox{\boldmath $L$}$.
We perturbe the system around the
de Sitter solution for $t=0$, i.e., $H(t =0) = h_i + \epsilon (0)$
and take $\dot{H}(t=0) = \dot{\epsilon}(0)$, as initial condition. 

\subsection{Saddle points and nodes}
In the neighborhood of these points the discriminant $\Delta$  is positive
and the eigenvalues $\lambda _{\pm}$ are real and different. For 
$det(\mbox{\boldmath
$L$})<0$ one has a saddle point, and for $det(\mbox{\boldmath
$L$})>0$ a node. The solution of (\ref{eq:equa}) is
\[
h = c_1 e^{\lambda _+ t} + c_2 e^{\lambda _- t},
\]
with 
\[
c_1 =-\frac{\dot{\epsilon}(0)-\epsilon (0)\lambda _-}{\lambda _-
-\lambda _+} 
\]
\[
 c_2 =\frac{\dot{\epsilon}(0)-\epsilon (0)\lambda _+}{\lambda _-
-\lambda _+}. 
\]
Upon integration one has for the scale factor
\begin{eqnarray}
R(t)\propto e^{h_i t}\exp \left [ \frac{c_1}{\lambda _+} e^{\lambda _+
t}+\frac{c_2}{\lambda _-} 
e^{\lambda _-t}\right],
\label{eq:R1}
\end{eqnarray}
\noindent which shows superinflationary expansion if initial
conditions are taken such that $c_{1},\ c_{2}$ are positive.
This type of evolution for $R(t)$ on time  has been obtained previously
in a different context\cite{vsiete}. In this case the fluid when submited to a 
small
perturbation, departs from equilibrium, expands much more 
rapidly than the de Sitter's.
In the case of nodes, and when $\lambda_+ +\lambda_- =  Q_X$ is positive
(negative), the node will be
unstable (stable).

\subsection{Attractors and repellors}
We study here the behavior of the scale factor near a sink
(an asymptotically stable attractor)  and a source (an asympotically 
unstable repellor). The solutions of the characteristic 
equation are complex quantities 
\[
\lambda _{\pm} = \frac{Q_X \pm i\sqrt{\mid 
\Delta \mid}}{2}
\]
where $\Delta = Q_X^2 + 4Q_H$. Then the solution of (\ref{eq:equa})
is 
\[
h =c_1 e^{\frac{Q_X}{2}t} \sin \frac{\sqrt{\mid \Delta \mid}}{2}
(t+c_2). 
\]
The integration constants can be determined through the initial
conditions. They read

\[
c_2 =  \frac{2}{\sqrt{\mid \Delta \mid}}\tan
^{-1}\left[\frac{\epsilon (0)\sqrt{\mid \Delta
\mid}}{2\dot{\epsilon}(0)-\epsilon (0) Q_x} 
\right ]
\] 
and 
\[
c_1 = \frac{\epsilon (0)}{\sin\frac{\sqrt{\mid \Delta \mid}}{2}c_2}. 
\]
Integrating the equation for $h$ one follows
\begin{eqnarray}
& &R(t) \propto e^{h_i t} \exp \left [\frac{2c_1}{Q_x^2+\mid \Delta \mid}\left 
(Q_x\sin \frac{\sqrt{\mid \Delta \mid}}{2}(t + c_2) \right. \right.\cr 
&-&\left. \left.  \sqrt{\mid \Delta \mid}
\cos \frac{\sqrt{\mid \Delta \mid}}{2}(t+c_2 )\right)\right ].
\label{eq:fac2}
\end{eqnarray}
For $Q_X<0$ we have an  attractor (the scale factor undergoes an oscillatory
approach to the de Sitter solution) and for $Q_X>0$ it is a source,
i.e., the scale factor deviates from the de Sitter solution.

\subsection{Degenerate nodes}
In this case $\Delta =0$ and $\lambda _+ = \lambda _- =\lambda =
\frac{Q_X}{2}$. The
solution of (\ref{eq:equa}) is 
\[
h= c_1 e^{\lambda t}+ c_2 t e^{\lambda t}. 
\]
Because of the initial conditions the integration constants are
\[
c_1 = \epsilon (0),\ \ \mbox{and}\ \ c_2 = \dot{\epsilon}(0)-\lambda
\epsilon (0).
\]
Integration of the expression for $h$ leads to
\begin{eqnarray}
R(t) \sim e^{h_i t} \exp \left [\frac{e^{\lambda t}}{\lambda }
\left(\epsilon (0) +
(\dot{\epsilon}(0)-\lambda \epsilon)(t-\frac{1}{\lambda})\right )\right ],
\label{eq:fac3}
\end{eqnarray}
hence $R(t)$ approaches to or separates from the de Sitter solution depending
on the sign of $\lambda$. The rate of evolution is faster than in the de Sitter
case.

\subsection{Energy conditions}
The weak energy condition (WEC) states that
$T_{ab}W^aW^b\geq 0$, where $T_{ab}$ is the energy-momentum tensor
given by (\ref{eq:Tab}) and $W^a$ a generic timelike vector. In our model
this condition reduces to
\begin{eqnarray}
\rho +\Lambda  \geq 0 .
\label{eq:wec}
\end{eqnarray}
The dominant energy condition (DEC) imposes $T_{ab}W^aW^b\geq 0$ and 
$-T^{ab}W_a$ to be
a non-spacelike vector which is equivalent to $T_{00}\geq\left|T_{ab}
\right|$\cite{vocho}. This conditions is fulfilled in our case only if
\begin{eqnarray}
-\rho\leq p +\Pi\leq \rho + 2\Lambda .
\label{eq:dec}
\end{eqnarray}
Finally, the strong energy condition (SEC) requires that
$T_{ab}W^aW^b+\frac{1}{2}T_a^a\geq 0$ which amounts to
\begin{eqnarray}
\rho - 2\Lambda +3p +3\Pi\geq 0.
\label{eq:sec}
\end{eqnarray}
These conditions can be rewritten in terms of $H$ and $\dot{H}$ as
\par
WEC: \hspace{1cm} $H^2\geq 0$\par
SEC: \hspace{1cm} $H^2+\dot{H}\leq 0$\par
DEC: \hspace{1cm} $\dot{H}\leq 0\;\; \mbox{and} \;\;
3H^2+\dot{H}\geq 0$\par
As occurs in the standard
inflationary scenarios the de Sitter solutions with $\Lambda \geq 0$
satisfy the WEC and DEC but not SEC.\\

\section{CONCLUSIONS}
We have carried out a detailed analysis on the stability of de Sitter
and static cosmological models, both in the truncated and full theory
for the viscous transport equation (with and without a cosmological constant).
We have shown that the conditions for the hydrodynamic approach and
the non-thermalizing condition in a de Sitter expansion can be
fulfilled 
simultaneously for sufficiently early times. When {\em no} cosmological
constant is considered the stability analysis for the de Sitter
solutions leads to similar results in all the cases,
i.e., the models are {\em unstable} only for $m>1/2$. It is remarkable that 
this result holds for both the truncated
and the full version of ET. On the other hand, when a {\em positive} 
cosmological constant is
included we see that the models can be stable for $m=1$ if $\Lambda$
is bounded from below.
It remains to be proved that this result holds for a generic $m>1/2$
other than $1$.\par
We have stressed the fact that for a radiation gas a different thermodynamic 
approach exists depending on whether the particle number density is
taken as an
independent variable or not. In the case of a power-law for the
temperature a relationship between $\gamma$ and $r$ exists.

\section*{Acknowledgements}
We would like to thank Roy Maartens, Diego Pav\'on and Winfried
Zimdahl for helpful discussions.\par
This work has been partially supported by the Spanish Ministry of
Education under Grant PB94-0718. One of us (J.T.) acknowledges financial 
support from a FPI grant.


\begin{references}
\bibitem[19]{dnueve}In fact it has been argued$^{13}$ that even
$T$ might rises in the de Sitter phase due to viscous dissipation
effects. It would increase the cross section $\sigma$ enhancing the inequality
in (\ref{eq:hydr}).
\end{references}
\end{document}